\documentclass[proceedings]{JHEP3}

\PrHEP{PrHEP hep2001}                   
\conference{International Europhysics Conference on HEP}                

\usepackage{epsfig}                   

\title{Limits on WIMP Dark Matter}

\author{\speaker{Neil Spooner} and 
                    Vitaly Kudryavtsev\\

         University of Sheffield, Dept. of Physics and Astronomy, 
         Hounsfield Rd. Sheffield, S3 7RH, UK\\ 
                            
        \email{n.spooner@sheffield.ac.uk},   \email{v.kudryavtsev@sheffield.ac.uk}}

\abstract{The current state searches for dark matter in the form of Weakly 
Interacting Massive Particles (WIMPs) using both direct and indirect
techniques is reviewed. Advances in recent years by various direct search experiments,
utilising technology able to record the nuclear recoil events expected
from elastic scattering by WIMPs, have allowed progress towards lower
limits to be made. In particular, the Edelweiss and CDMS collaborations are 
achieving sensitiviy able to challenge data from DAMA interpreted as 
evidence for WIMPs of mass in the region of 60 GeV. Meanwhile, indirect 
searches, based on observing the annihilation products of 
neutralino-neutralino interactions in the Earth, Sun and Galaxy, have 
produced intriguing results.  For instance, analysis by Superkamiokande 
now suggests limits comparable with the best direct search results. }

\begin{document}

  \section{WIMP Direct Searches} 

{WIMPs interact with normal matter by elastic scattering off nuclei.  The
energy deposited by the resulting recoil nuclei or atoms has a
characteristic exponential spectrum.  This is determined mainly by the
kinematics of the interaction, the WIMP mass relative to that of the
recoiling nuclei and the velocity of the WIMP, determined by the velocity
of the Earth through the galactic halo.  The favoured range of WIMP masses,
velocities and likely cross sections (for instance for MSSM) lead to recoil
spectra expected to have energy ranging from a few keV upto a few hundred
keV with rate $<$1 kg$^{-1}$day$^{-1}$.  The latter rate is typically
a factor of 10$^{6}$ lower than the ambient rate from background gammas due to
surrounding natural radioactivity~\cite{[1]}.

These characteristics determine basic requirements of direct detection
technology, the need for low energy threshold and some means of identifying
genuine recoils from the much higher rate of background electron recoils. 
The latter is feasible in principle because the energy loss per unit track length
(dE/dx) for electrons is typically 10 times lower than for nuclear
recoils~\cite{[2]}.  However, any neutrons present, such as those produced
by cosmic ray muons, can produce background nuclear recoils
indistinguishable from those expected from WIMP interactions.  Therefore,
it is essential also that direct WIMP searches be performed in deep
underground sites, typically $>$1000 mwe, where this flux is negligible or
can be sufficiently reduced using neutron shielding.

Several technologies hold out prospects for achieving the requirements
above but the most favoured at present are ionisation, scintillation and
low temperature bolometric devices.  Germanium ionisation detectors, used
initially for double beta decay searches, set the first limits.  More
recently of note has been the Heidelberg-Moscow detector and the HDMS
prototype Ge detectors~\cite{[3]} operating at Gran Sasso.  These have
produced currently competitive limits.  However, detectors using ionisation
alone have no means of actively distinguishing nuclear recoils from
electron background.  Hence only limits can be set, based on the measured
continuum background.  The recent development of Ge detectors have thus
tended to concentrate on material purification, in an effort to reduce
intrinsic radioactivity.  However, development towards larger mass Ge
(10s-100s kg), exemplified by the GENIUS project~\cite{[4]} and
others~\cite{[5]}, may allow observation of the expected annual modulation
of the dark matter event rate, arising from the earth's varying speed
through the Galaxy.

Scintillation and low temperature detectors provide a route to the required
additional information for recoil identification~\cite{[6],[7]}.  In the
former, in crystal scintillators or liquid noble gases, the high dE/dx for
nuclear recoils results in pulse decay times a factor of 0.3-0.5 shorter than for
electrons.  Statistical analysis can then be used to identify a population
of faster events~\cite{[8],[9]}.  First limits were set in 1994-5 using
this idea in NaI~\cite{[9],[10]}.  Subsequently, following improved
sensitivity, the UKDMC group at the Boulby site discovered a population of
fast events at low energy in NaI, possibly due to surface alpha particles
~\cite{[11],[12]}.  Meanwhile the Rome group (DAMA), operating 100 kg of
NaI at Gran Sasso, has reported an annual fluctuation in the total count
rate over 4 years.  They interpret this as consistent with the annual
fluctuation predicted for WIMPs~\cite{[13],[14]}.  However, this is not yet
widely accepted because the technique does not separate nuclear recoils
from the much larger low energy background which, in principle, could be
subject to other modulating systematics~\cite{[15]}.

Several experiments based on counting total events in low temperature
bolometers are underway and have set limits, notably by CRESST and the
Milan group~\cite{[16],[17],[18]}.  However, of greater significance are
schemes in which nuclear recoil identification is achieved in bolometric
detectors by combining with simultaneous observation of ionisation or
scintillation.  The former is used by the CDMS-I and Edelweiss experiments,
the latter is being developed by CRESST~\cite{[19],[20]}.  The CDMS
experiment, although not yet located deep underground and hence needing to
subtract neutron background, has presented data that appear to exclude the
Rome result~\cite{[21]}.  They reach a spin independent WIMP-nucleon limit
around $2 \times 10^{-6}$ pb in the mass range 20-100 GeV. Edelweiss have also
released results that significantly cut into the Rome allowed region but
with the advantage that no neutron subtraction is needed as they already
operate deep underground, at the Modane site~\cite{[22]}.  Recent results
are summarised in Fig. 1, reproduced from~\cite{[22]}.

\FIGURE{\epsfig{file=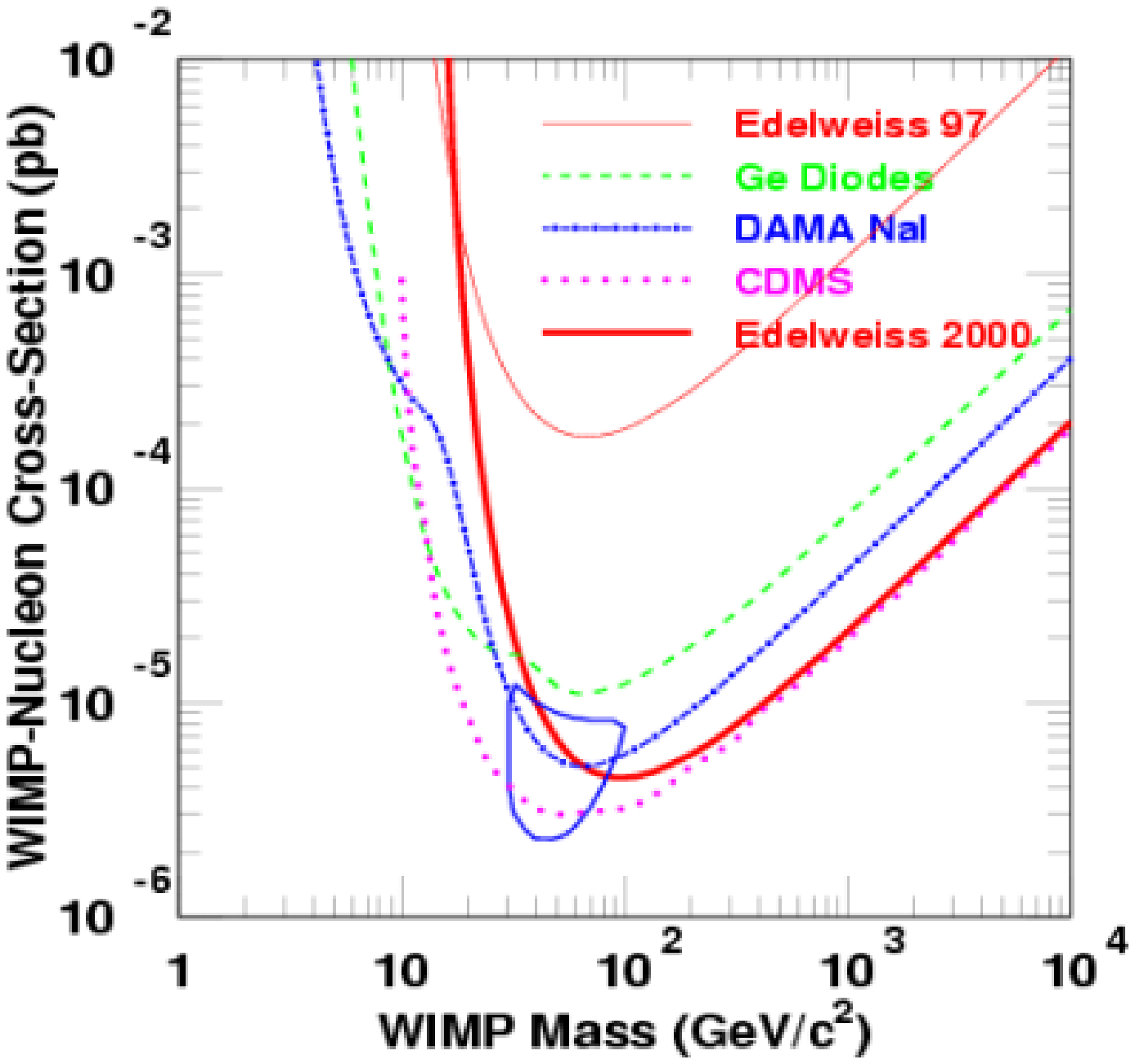,width=3.5in}
\caption{Recent results for spin independent WIMP searches.}
\label{fig:edel.eps}}
       
New generations of experiment are being developed now aimed at factors of 10-1000
sensitivity improvements over 2-5 years. CDMS-II will be an expansion of
the CDMS-I experiment to be run in the Soudan Mine.  CRESST-II, using
scintillation plus low temperature technology at Gran Sasso is predicted to
achieve similar sensitivity.  Notable also is the growing interest in
liquid Xe.  Early experiments by the Rome group~\cite{[23]} have now been
supplemented by a Japanese group in Kamioka~\cite{[24]} and a major effort
on xenon by the UKDM collaboration with UCLA, Torino, ITEP and
Columbia~\cite{[25]}.  This consortium is constructing a series of liquid
Xe experiments at Boulby.  ZEPLIN I, now running at Boulby, is based on
pulse shape discrimination.  ZEPLIN II (see Fig.  2) makes use of
simultaneous collection of scintillation and charge to achieve factors of
10-100 improved sensitivity and ZEPLIN III incorporates a high E-field in
the liquid to enhance the recoil signal.

\FIGURE{\epsfig{file=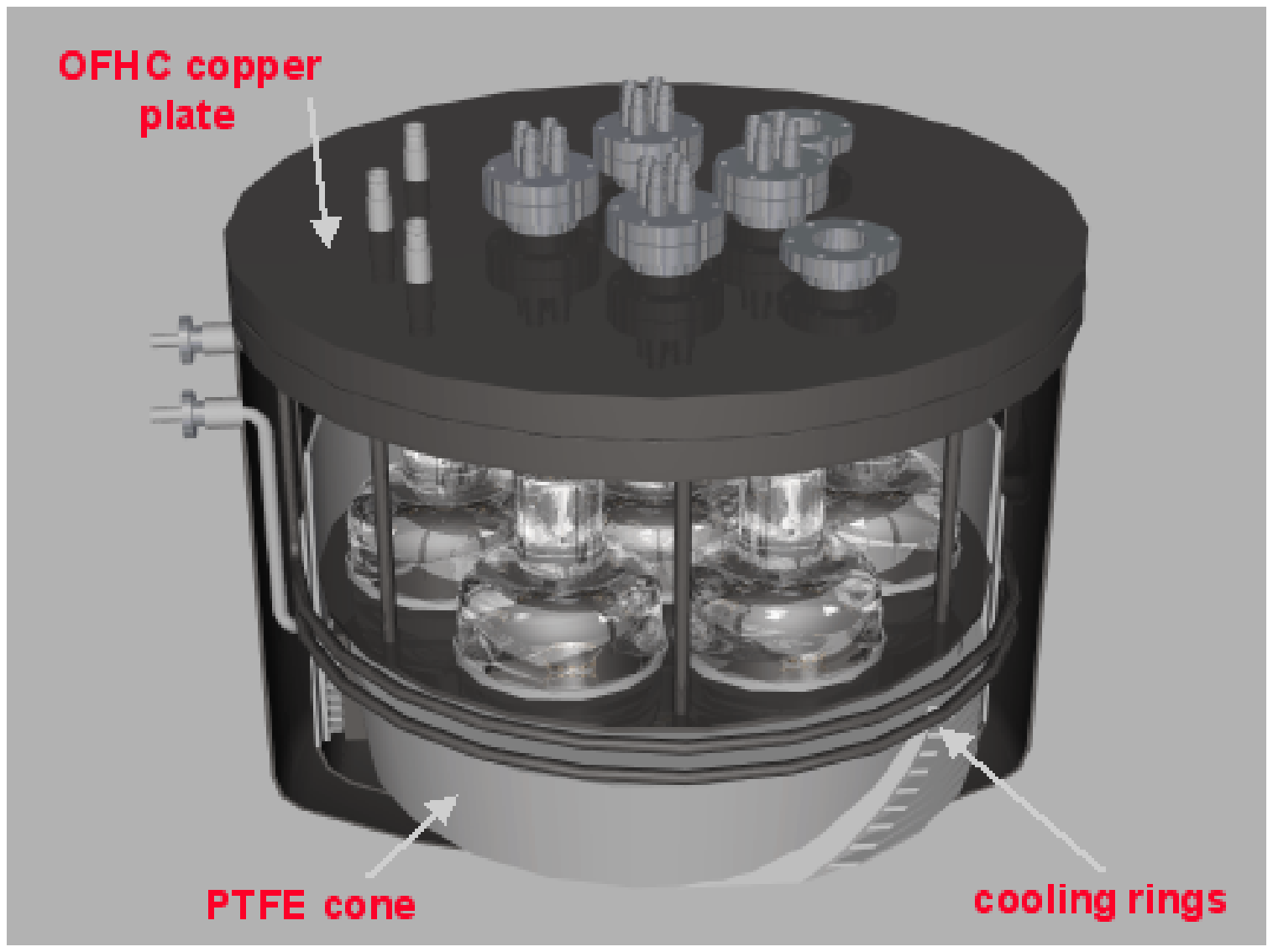,width=3.5in}
\caption{The ZEPLIN II liquid Xe detector of UKDM/UCLA/Torino.}
\label{fig:ZII.eps}}

A 1000 kg liquid Xe detector, ZEPLIN-MAX, is currently being designed by
the UKDM to achieve sensitivity below $<$10$^{-9}$ pb.  Fig.  3 ilustrates
the potential sensitivity of the liquid xenon experiments, estimated from
preliminary results with ZEPLIN I.

\FIGURE{\epsfig{file=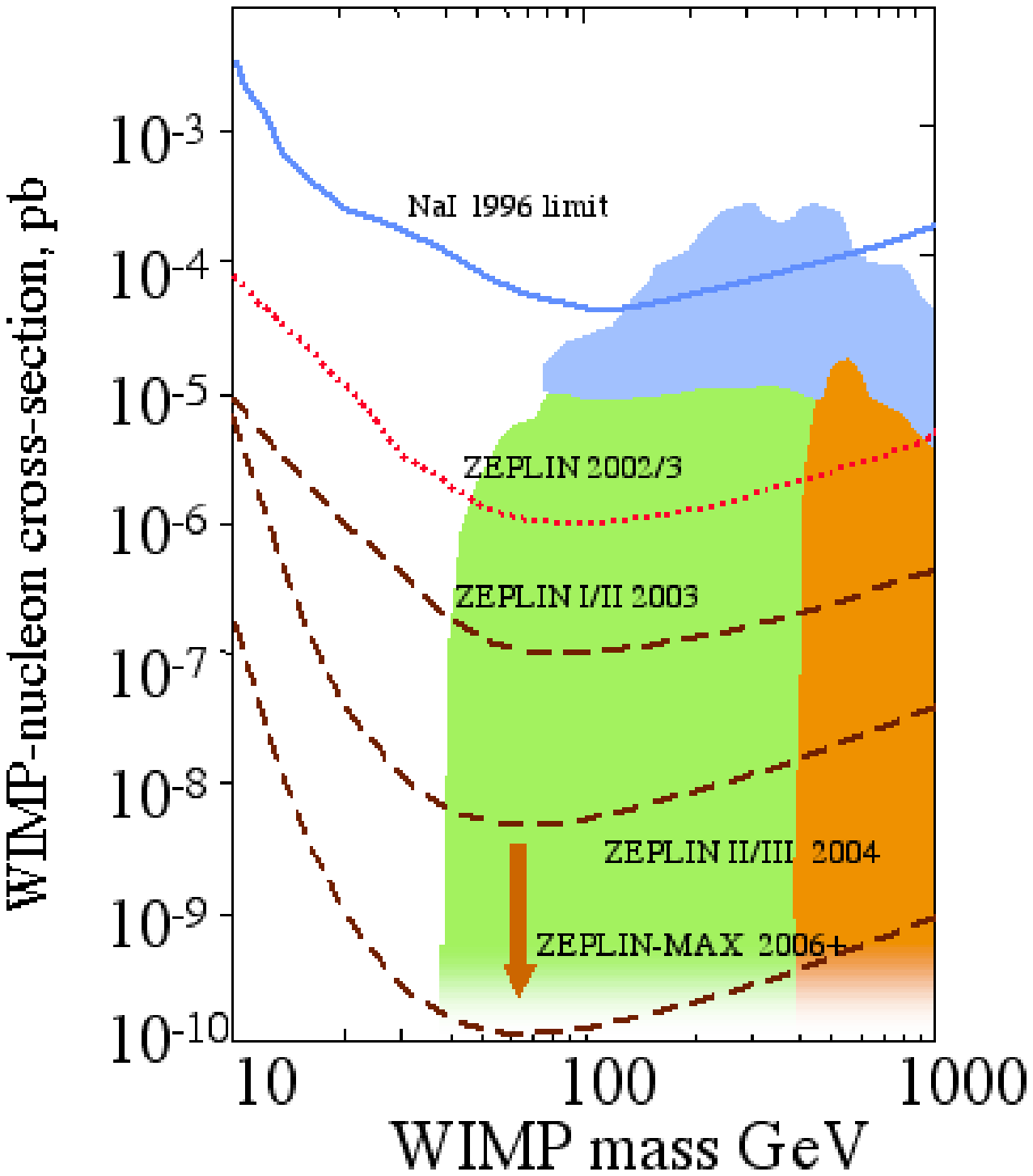,width=3.5in}
\caption{Predicted sensitivity of the ZEPLIN series of xenon dark matter detectors.} 
\label{fig:ZEPLIM.eps}}

Other novel techniques, in particular using superheated droplet detectors,
may also eventually prove very sensitive~\cite{[26],[27]} but ultimately
the most convincing demonstration of the existence of WIMPs would be
correlation of the direction of nuclear recoils with our motion through the
Galaxy.  The most promising technique to achieve this is by means of a low
pressure Time Projection Chamber in which recoil tracks of a few mm length
can be imaged.  A UK/US collaboration is now running such a device of 1
m$^{3}$ called DRIFT-I at Boulby~\cite{[28],[29]}. Fig. 4 shows
preliminary underground Cf neutron calibration data for a short 43 min run
from this detector taken with no passive shielding.  Events are plotted as
number of ionizing pairs (NIPs) versus a discrimination parameter R2 that
quantifies track range.  Gammas are expected to show as events with
relatively high R2, close to the vertical axis.  However, sensitivity to
gammas is so low that only neutron (nuclear recoil) events are observed,
confirmed by runs without the neutron source.  Such a directional dark
matter detector offers the prospect of a dark matter "telescope" able to
distinguish possible different velocity components of the dark matter that
have been suggested could exist~\cite{[30]}.

\FIGURE{\epsfig{file=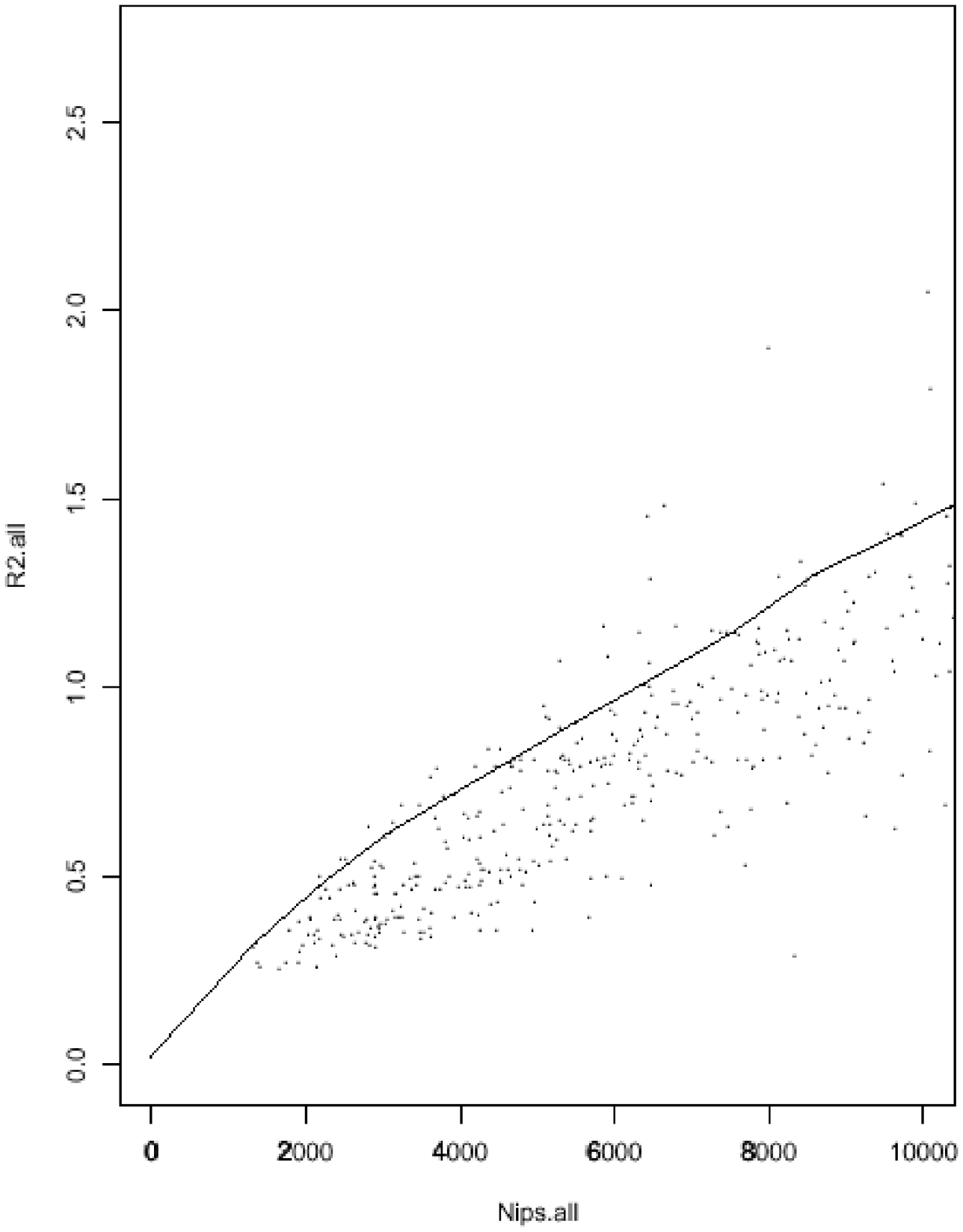,width=3in}
\caption{Neutron detection by the DRIFT-I directional dark matter detector.}
\label{fig:DRIFTn.eps}}

  \section{WIMP Indirect Searches} 

{If WIMPs are Majorana neutralinos then pair annihilations can occur and it
may be possible to detect the resulting neutrinos, gamma rays, positrons or
antiprotons.  Such indirect detection of WIMPs is quite complementary to
direct observation though much more model dependent, affected for instance
by possible non-maxwellian velocity components in the halo.  Indirect
searches can be more sensitive to high mass WIMPs while neutralino models
which produce low direct detection rates can sometimes produce substantial
annihilation rates~\cite{[31]}, for example through the gamma-gamma
channel.  The most likely scenario is to search for high energy
neutrino signals from the Sun, Earth, or galactic centre where the WIMP
density may be sufficiently enhanced by gravitational capture.  The halo
may provide a further source if the dark matter is clumpy~\cite{[32]}. 
Neutrinos, like annihilation gammas, have the advantage of maintaining
their original direction.

Observation of muon neutrinos provide the best hope for the neutrino
channel since the resulting upgoing muons produced in the Earth can be
distinguished from background down-going atmospheric muons and have long
range in present and planned Cherenkov neutrino detectors.  These include
AMANDA, ANTARES, IceCube, Baikal and NESTOR (see, for example, \cite{[34],[35],[35a]};
see also \cite{[33]} for a recent review).  The
Sun, being dominated by hydrogen, is particularly favourable, with
predictions of the muon rates for different neutralino models also easier
to calculate.  Nevertheless, calculations have been performed for both Sun
and Earth~\cite{[36]}.

Present neutrino experiments have already provided significant limits on
the Sun and Earth muon flux sufficient to constrain MSSM
models~\cite{[37],[38]}.  Limits in the range 10$^{3}$-10$^{4}$ muons
km$^{-2}$yr$^{-1}$ are found for the Sun above 10$^{2}$ GeV and down to
10$^{3}$ muons km$^{-2}$yr$^{-1}$ above 10$^{3}$ GeV for the
Earth (see, for example, \cite{[39a]} and references therein; see also
\cite{[39]} for a review).  The latter is sufficient to indicate a possible
contradiction with the DAMA direct search signal.  A recent analysis by the
SuperK collaboration to produce a WIMP-nucleon cross section limit using
combined Sun, Earth and galactic centre data (see Fig.  5) also appears to
exclude parts of the DAMA allowed region~\cite{[40]}. The AMANDA and
ANTARES experiments are now aiming for km$^{2}$ experiments that would
provide a factor of 10$^{4}$ improvement in sensitivity, sufficient to test large
parts of the MSSM parameter space~\cite{[41]}.

\FIGURE{\epsfig{file=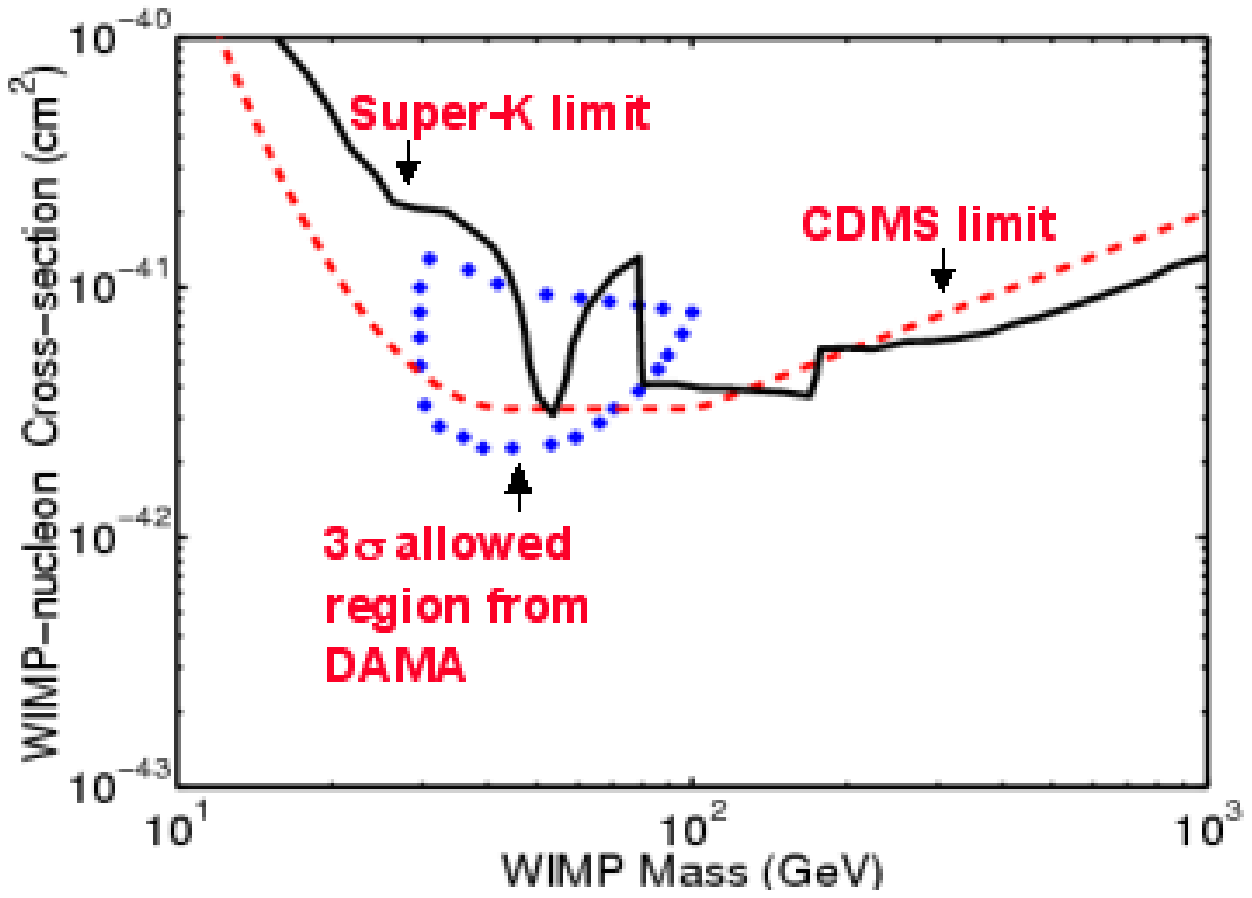,width=3.5in}
\caption{Combined Sun, Earth and galactic centre limit from SuperK~\cite{[40]}.}
\label{fig:SuperK.eps}}

Searches for antiproton, positron and gamma ray lines from annihilation in
the halo are also underway.  The former two channels are hindered by
uncertainty in galactic propagation models and the featureless nature of
predicted spectra.  Nevertheless balloon borne experiments to search for
neutralino annihilation antiprotons at the top of the atmosphere have been
performed, for instance by Bess and Caprice~\cite{[42],[43],[44]}, to be
compared with predictions of secondary antiproton background~\cite{[45]}. 
The space experiment AMS aims also to undertake a search~\cite{[46]}. 
Despite large possible systematic effects, such as from cosmic-ray induced
antiprotons, interesting limits can be placed for the highest annihilation
rates~\cite{[47]}.  Balloon observation of the positron continuum have also
been performed. No excess over predictions from secondary positron production
has been observed so far \cite{[48],[49]}.

Although very sensitive to the local neutralino halo density, annihilation
gamma-ray lines from the halo can be observed in principle by existing or
planned Air Cherenkov Telescopes (ACTs) such as Veritas, Whipple, STACEE,
CELESTE, MAGIC and MILAGRO, or by space-borne detectors including EGRET and
GLAST. Indeed this technique may be the only one available to probe for
heavy (TeV) stable neutralinos. The ACTs have acceptance angles suitable
for searches of possible galactic centre signals.  The high energy
resolution of GLAST makes it suitable for high precision line searches. 
Recent MSSM calculations show that for a "standard" halo, for instance,
Veritas and GLAST have discovery potential, with TeV masses
accessible~\cite{[50],[51]}.

\end{document}